\documentclass[twoside]{dis07}
\usepackage[latin1]{inputenc}
\usepackage[dvips]{graphicx,epsfig,color}
\usepackage{wrapfig,rotating}
\usepackage{amssymb,amsmath,array}

\begin{document}
\title{Parton shower Monte Carlos vs resummed calculations for interjet energy flow observables.}

%***********************************************************************
% AUTHORS INFORMATION AREA
%***********************************************************************
\author{Mrinal Dasgupta
%
% Optional short acknowledgment: remove next line if non-needed
\thanks{Work in collaboration with Andrea Banfi and Gennaro Corcella.}
%
% DO NOT MODIFY THE FOLLOWING '\vspace' ARGUMENT
\vspace{.3cm}\\
%
%Addresses and institutions (remove "1- " in case of a single institution)
School of Physics and Astronomy, University of Manchester  \\
Schuster Building, Brunswick street, Manchester M13 9PL, United Kingdom.}
%
% Remove the next three lines in case of a single institution

%***********************************************************************
% END OF AUTHORS INFORMATION AREA
%***********************************************************************

\maketitle \vspace{-5.0cm} \hfill MAN/HEP/2007/12 \vspace{4.55cm}

\begin{abstract}
Parton showers in Monte Carlo event generators reflect to a certain accuracy our understanding of QCD radiation at all orders. For observables sensitive to 
interjet energy flow in well defined regions of phase space, 
it has been known for some time that relevant all-order dynamics is substantially more complicated than that encoded via angular ordering 
in parton shower algorithms,  even to leading logarithmic accuracy. Here we investigate the extent of the numerical mismatch between leading logarithmic analytical estimates (resummation) and parton showers in an effort to better understand the accuracy of parton showers for such observables.
\end{abstract}

\section{Introduction}
Monte Carlo (MC) event generators are the most important tools at our disposal in the search for new 
physics beyond the standard model. At the large hadron collider (LHC) for example, alongside the hard process where the production of hitherto unseen particles is expected to occur, there will in general 
be copious amounts of initial and final state QCD radiation, conversion of partons into hadrons (hadronisation), effects due to spectator interactions, multiple hard scatterings and pile-up. 

In such a complex hadronic environment  as good a  modeling as possible of all of the above physics aspects, is of clear benefit in order to exploit the full discovery potential of the LHC.
MC tools exist and are being further developed to provide a simulation of the physics described above.
However of the whole range of effects mentioned above as accompanying the hard process, 
the only part that we understand directly from first principles is 
perturbative QCD radiation. This understanding is in principle 
what ought to be reflected in the 
parton shower algorithms in various event generators.  

Another source of information on QCD at all orders is provided by 
analytical tools such as those employed in resummation 
of large logarithms in QCD.
This field too has seen some interesting 
relatively recent developments with the 
observation that for a large class of observables (those measuring flow of hadronic energy into limited regions of phase space and hence called 
non global observables) relevant all-order dynamics is much more complicated 
than previously imagined \cite{DassalNG1,DassalNG2}. This more 
complex soft gluon dynamics is not fully described by parton showers in event generators such as HERWIG \cite{hw} and PYTHIA \cite{py} \footnote{This dynamics is in fact included only in the ARIADNE event generator \cite{leif}}. 
The HERWIG parton shower for example is based on angular ordering of soft 
gluon radiation which correctly captures the leading logarithms for several observables. 
However for non global observables the angular ordering approximation is not formally correct at single logarithmic accuracy, which in some important cases is also leading logarithmic accuracy. In these cases the HERWIG shower cannot be a priori expected to describe even
the leading logarithmic perturbative behaviour. This is even more the case for the old 
PYTHIA shower (before v 6.3) where angular ordering is imposed on showers pre-ordered in invariant mass, which leads to a poor decription of soft emissions at large angles and consequently non-global observables in particular.
This is potentially worrying since non-global observables such as interjet energy flow are commonly measured experimentally and moreover they are used to 
improve MC models by tuning the model parameters to data on such observables e.g. hadronic activity in  {\it{toward}} or {\it{away}} regions wrt the 
leading jet, in hadron collisions. 

Given the prevalence and importance of non-global observables in QCD studies, it is worth investigating 
quantitatively how good a description is provided, in such cases, 
by the different parton 
showers e.g. those in HERWIG and PYTHIA. This can be done by comparing the parton showers to the state-of the-art analytical resummations in an effort to see 
what 
perturbative 
accuracy can be relied upon in the corresponding MC descriptions. 

In order to get a clearer idea about what one may expect from parton showers such as those ordered in angle we first study to what extent 
the leading single-logarithmic behaviour can be obtained by assuming angular 
ordering for energy flow into a rapidity interval between hard jets. 
Thus we compare the full single-logarithmic resummation (possible only in the large $N_c$ limit) with a numerical result obtained by addressing just angular ordered configurations. We then check if our conclusions are borne out by comparing the resummation to the real HERWIG parton shower. We then compare the results to the PYTHIA event generator both with the old shower and the new shower and draw some conclusions.

\section{Angular ordering and non global observables}
The observable we shall study is the differential $E_t$ flow, $\frac{1}{\sigma}\frac{d\sigma}{dE_t}$ into a rapidity interval of width $\Delta \eta$ between a pair of jets. For simplicity we shall take these to be back-to--back jets produced in $e^{+}e^{-}$ annihilation but our conclusions generalise straightforwardly to gaps between jets in any hard process. This observable involves other than the scale $E_t$ an additional scale $Q$, the hard scale for the process. Typically $E_t \ll Q$ and one then has to deal with logarithms $\alpha_s^n \ln^m Q/E_t$, with $m \leq n$. We note that the leading logarithms in this case are single logarithms i.e those with $m=n$. 

The resummation of the leading logarithms has been possible only in the large $N_c$ limit \cite{DassalNG1,DassalNG2}. In this limit the complex colour algebra involved in the soft large-angle multi-gluon emission probabilities is greatly simplified and can be encoded in a dipole evolution Monte-Carlo algorithm, since the emitting ensemble reduces to a system of dipoles. Considering a dipole configuration 
$C$ we can write the probability of 
emitting a soft gluon at scale $L' = \ln Q/\omega'$, where $\omega'$ is the gluon energy, as :
\begin{equation}
P_C'(L') = \alpha_s(L') \Delta_C(L',L) F_C(\theta'\phi') P_C(L),
\end{equation}
where $C'$ is the new configuration of dipoles after the emission of the gluon, $\Delta(L',L)$ is the no-emission probability between scales $L$ and $L'$ and $F_c(\theta',\phi')$ is the angular part of the dipole real emission probability.
Thus explicitly we have 
\begin{equation}
F_C(\theta_k,\phi_k) = \sum_{\mathrm{dipoles-ij}} \frac{2 C_A\left(1-\cos\theta_{ij}\right)}{(1-\cos\theta_{ik}(1-\cos\theta_{jk})}
\end{equation}
Then the result that contains the resummation of the leading single-logarithms 
is a form factor, $\Sigma$, which can be written
\begin{equation}
\Sigma(\alpha_s L) = \sum_{C|\Omega_{empty}}P_c(L), \, L= \ln Q/E_t.
\end{equation}
In other words the resummed result for the probability that the transverse energy in the gap is less than some value $E_t$ is obtained by summing over all dipole configurations such that the gap region $\Omega$ is free from emissions above that scale.
In order to study the impact of angular ordering 
we replace $F$ by an angular ordered (AO) approximation \footnote{This replacement would be exactly correct 
for global observables where one could freely integrate over azimuthal angles.}:
\begin{equation}
F(\theta_k,\phi_k) \to \frac{\left(\cos \theta_{ik}-\cos \theta_{ij}\right)}{\left(1-\cos\theta_{ik}\right)}+ \frac{\left(\cos \theta_{jk}-\cos \theta_{ij}\right)}{\left(1-\cos\theta_{jk}\right)}
\end{equation}
such that the emission $k$ is produced in cones around the hard legs $i$ and $j$  with angular size set by the opening angle of dipole $ij$.
This angular ordering is expected to correspond to the piece included in e.g. the HERWIG shower.

\begin{wrapfigure}{r}{0.5\columnwidth}
\centerline{\includegraphics[width=0.45\columnwidth]{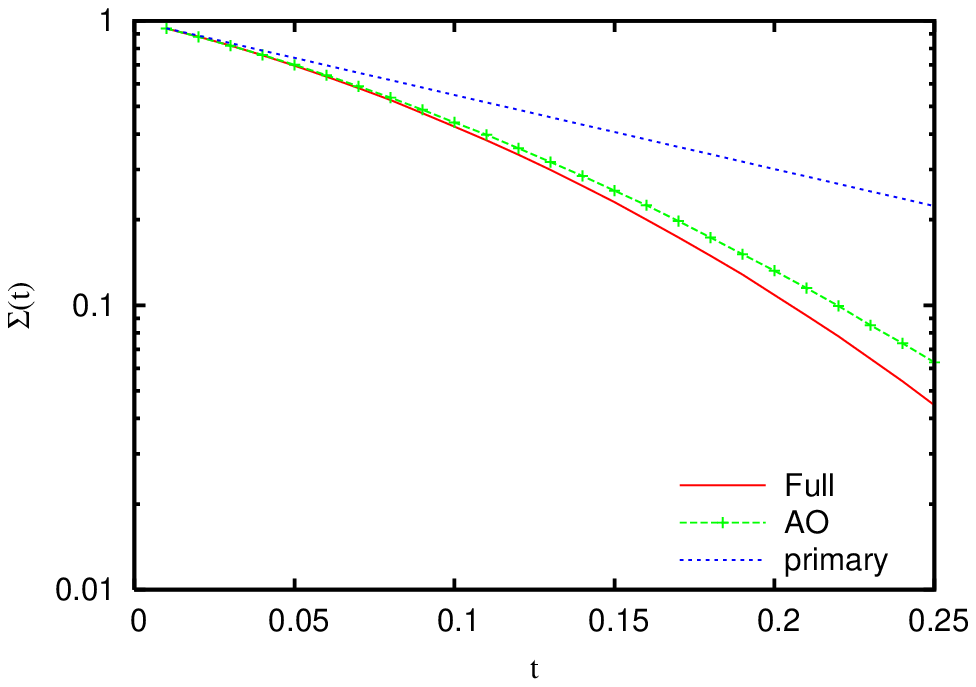}}
\caption{}
\label{Fig:slice}
\end{wrapfigure}

In Fig.~\ref{Fig:slice},we plot the resummed integrated $E_t$ cross-section (equivalently the form factor $\Sigma$) 
as a function of $t \sim \frac{\alpha_s}{2\pi} \ln \frac{Q}{E_t}$ in both the AO approximation as well as the full 
resummation in the large $N_c$ limit, for $\Delta \eta =1.0$. 
We note that angular ordering has only a modest effect and the results at say $t=0.15$ corresponding to $Q=100$ GeV $E_t =1$ GeV 
differ by just 10 \%. Also shown there (the 'primary' curve) is the resummation in the independent emission approximation assuming soft gluons to be emitted by just the 
hard partons with no correlated gluon emission other than effects included in 
the running coupling. The angular ordered and full results differ 
substantially from this stronger assumption.

\section{Comparison to HERWIG and PYTHIA}
\begin{wrapfigure}{r}{0.5\columnwidth}
\centerline{\includegraphics[width=0.45\columnwidth]{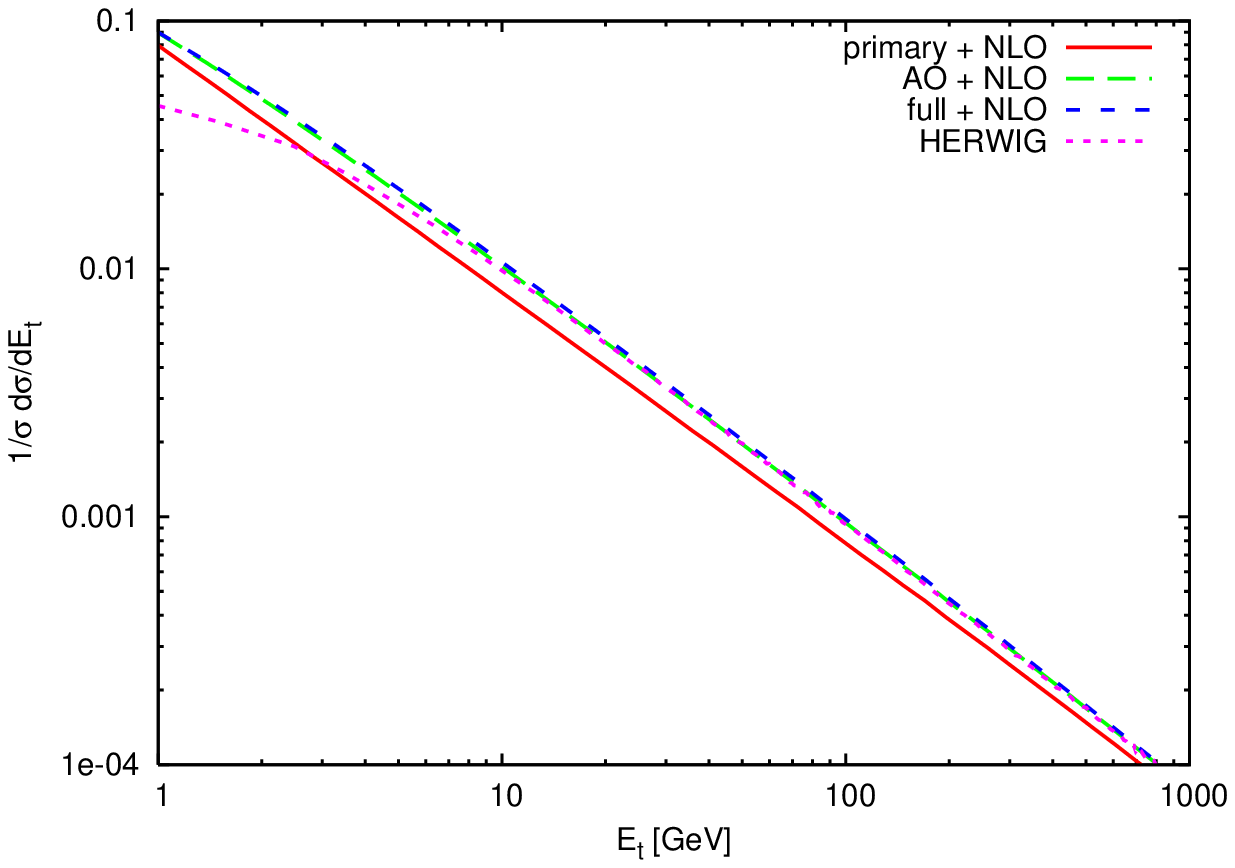}}
\caption{}
\label{Fig:hwslice}
\end{wrapfigure}
We now proceed to compare to actual parton showers in HERWIG and PYTHIA. Since we are interested in the dependence on the logarithmic variable $t$ we can in fact go to very large $Q$ values, which helps us to kill potentially spurious 
subleading terms of relative order $\alpha_s(Q)$, that would be present in the MC showers. Once we draw conclusions for a given value of $t$ , we can 
apply those observations to an appropriate $E_t$ value for {it{any}} $Q$.
Given that angular ordering produces results that are in reasonable agreement with resummation we can a priori expect the HERWIG shower to also not differ substantially. 
This is in fact the case as can be seen from Fig.~\ref{Fig:hwslice} where once again for $t=0.15$ we note just a 10 \% difference between HERWIG and resummation (essentially indistinguishable curves on the plot here). 

\begin{wrapfigure}{r}{0.5\columnwidth}
\centerline{\includegraphics[width=0.45\columnwidth]{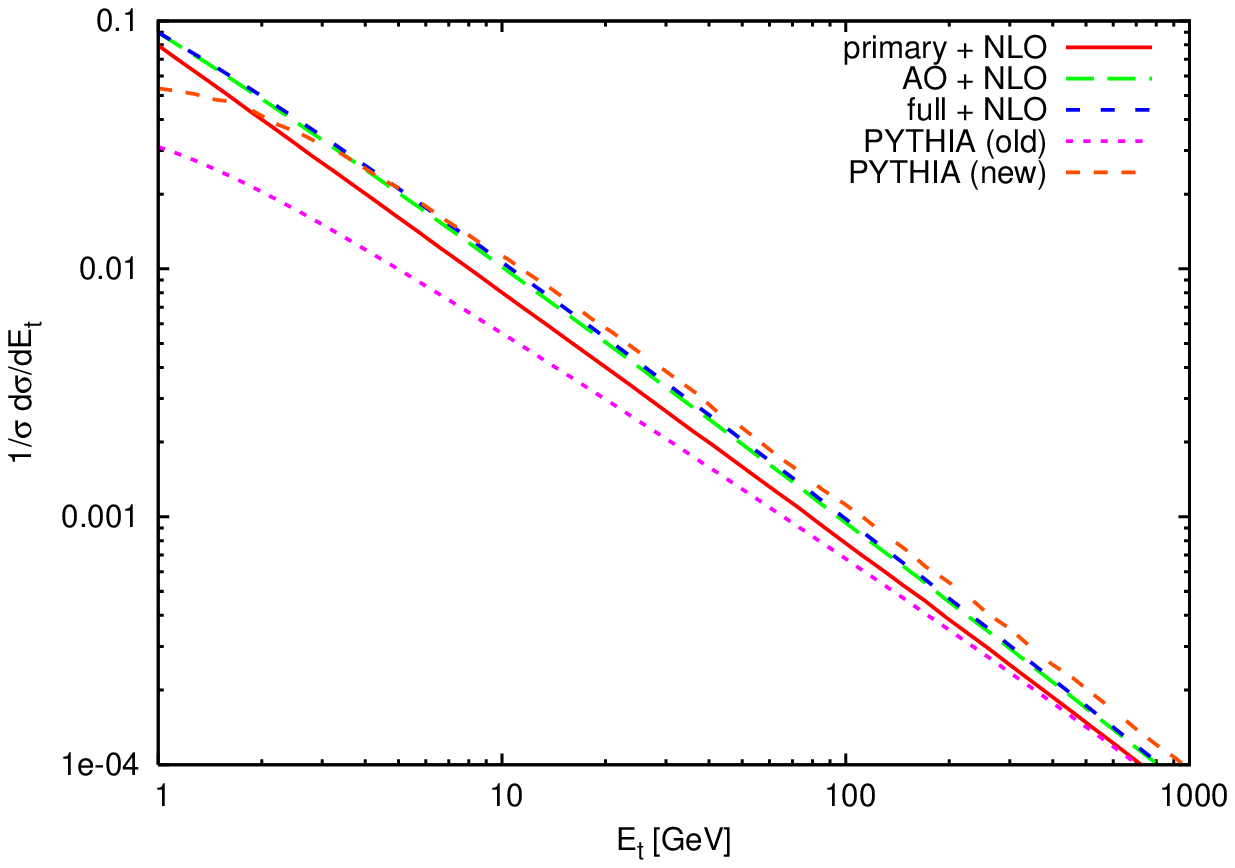}}
\caption{}
\label{Fig:pyslice}
\end{wrapfigure}
For the old PYTHIA shower where angular ordering is imposed on top of ordering in invariant mass the results are significantly different.
Here one notes from Fig.~\ref{Fig:pyslice} a discrepancy of 50 \% for $t=0.15$ which may be misattributed to other (non-perturbative) physics 
during the procedure of tuning PYTHIA to data. Bearing in mind that the non global logarithmic effects are not universal and are absent for other (global) observables this is a potentially dangerous and inaccurate way to account for them.

For the new PYTHIA shower which implements angular ordering of soft radiation via a dipole type phase-space 
the results show a marked improvement and a difference of only 
7.5 \% is seen at $t=0.15$. However as we increase the width of the rapidity slice in question the description from the new PYTHIA shower deteriorates and we observed that for $\Delta \eta =3.0$ the new shower performs poorly when compared to resummation and HERWIG results.

\section{Conclusions}
We can conclude from our studies that while angular ordering as implemented in QCD parton showers does not formally give the correct single 
logarithmic behaviour for non-global observables, numerically one does not have to worry too much about the difference between angular ordered and resummed results even in this case. However ordering in other variables such as invariant mass and angle (PYTHIA old) gives rise to a poor perturbative description. This incorrect perturbative estimate may be adjusted for in other non-perturbative aspects of the MC model but this is not optimally accurate. We also observed that the new PYTHIA shower is substantially better than its earlier counterpart but noted a puzzling discrepancy at large values of the interjet rapidity interval $\Delta \eta$ \cite{BanDasCor}, which we hope can be clarified in the near future.
 
It is imperative in our opinion, especially for observables related to  energy flow between hard jets, to compare the results from HERWIG and PYTHIA both at the parton and hadron level. If large discrepancies are observed at parton level 
information from analytical resummed calculations where available could be exploited to better understand the uncertainty and limitations of the MC 
descriptions.

% ****************************************************************************
% BIBLIOGRAPHY AREA
% ****************************************************************************

\begin{footnotesize}
% IF YOU DO NOT USE BIBTEX, USE THE FOLLOWING SAMPLE SCHEME FOR THE REFERENCES
% ----------------------------------------------------------------------------

% ----------------------------------------------------------------------------

% IF YOU USE BIBTEX,
% - DELETE THE TEXT BETWEEN THE TWO ABOVE DASHED LINES
% - UNCOMMENT THE NEXT TWO LINES AND REPLACE 'Name_Of_Your_BibFile'

%\bibliographystyle{unsrt}
%\bibliography{Name_Of_Your_BibFile}
% example of Name_Of_Your_BibFile.bib
% @Article{Turcato:2006ch,
%      author    = "Turcato, M.",
%  collaboration = "ZEUS and H1",
%      title     = "Lepton flavour violation and charmonium physics at HERA",
%      journal   = "Nucl. Phys. Proc. Suppl.",
%      volume    = "162",
%      year      = "2006", 
%      pages     = "283-287",
%      SLACcitation  = "%%CITATION = NUPHZ,162,283;%%"
% }
% 
% @Unpublished{Gogitidze:2007du,
%      author    = "Gogitidze, N.",
%  collaboration = "H1", 
%      title     = "Prompt photons and particle momentum distributions at
%                   HERA", 
%      year      = "2007",
%      note    = "hep-ex/0701033",
%      SLACcitation  = "%%CITATION = HEP-EX 0701033;%%"
% }

\end{footnotesize}

% ****************************************************************************
% END OF BIBLIOGRAPHY AREA
% ****************************************************************************

\end{document}